\documentclass[10pt]{iopart}
\pdfoutput=1
\usepackage{graphicx}
\usepackage{xcolor}
\eqnobysec

\newcommand{\bea}{\begin{eqnarray*}}
\newcommand{\beq}{\begin{equation}}
\newcommand{\beqa}{\begin{eqnarray}}
\newcommand{\eea}{\end{eqnarray*}}
\newcommand{\eeq}{\end{equation}}
\newcommand{\eeqa}{\end{eqnarray}}

\renewcommand{\d}{{\rm d}}
\def\e{{\rm e}}	

\newcommand{\eq}{{\rm eq}}

\newcommand{\eps}{\varepsilon_t}
\renewcommand{\u}{{\bar u_t}}

\newcommand{\erfc}{\mathop{\rm erfc}}
\newcommand{\typ}{{\rm typ}}
\newcommand{\g}{g}

\newcommand{\s}{{\sigma}}
\newcommand{\gam}{{\gamma}}
\newcommand{\comport}[3]{\mathrel{\mathop{#1}\limits_{#2}^{#3}}}
\newcommand{\prob}{{\cal P}}
\renewcommand{\l}{\ell}

\begin{document}
\title{Coarsening dynamics of zero-range processes}
\author{Claude Godr\`eche and Jean-Michel Drouffe}

\address{Institut de Physique Th\'eorique, Universit\'e Paris-Saclay,
CEA and CNRS, \\ 91191 Gif-sur-Yvette, France}

\begin{abstract}
We consider a class of zero-range processes exhibiting a condensation
transition in the stationary state,
with a critical single-site distribution
decaying faster than a power law.
We present the analytical study of the coarsening dynamics 
of the system on the complete graph, both at criticality and in the condensed phase.
In contrast with the class of zero-range processes with critical single-site distribution decaying as a power law,
in the present case
the role of finite-time corrections are essential for the understanding of the approach to scaling.
\end{abstract}


\section{Introduction}

While the static properties of zero-range processes (ZRP) are by now well understood~\cite{spitz,andj,hanney,lux}, dynamical properties are far less investigated.
This is all the more true when the model exhibits a condensation transition in the stationary state.
In such instances, of particular interest is the long-time evolution of the system starting from a homogeneous disordered initial condition.
In the scaling regime, a coarsening phenomenon takes place, i.e., a group of sites, whose number decreases, progressively become more populated, a process followed in the late-time regime by the appearance of a condensate. 
A few studies have been devoted in the past to the coarsening dynamics of the class of ZRP with a single-site critical distribution decaying as a power law~\cite{zeta1,zeta2,cg2003,gss}.
These studies yield analytical results when the dynamics takes place on the complete graph and in the thermodynamical limit~\cite{zeta1,zeta2,cg2003}.
In contrast, the knowledge of the dynamics of the one-dimensional system only relies on numerical 
work and heuristic arguments \cite{cg2003,gss}. 
When the density is just equal to the critical density one rather speaks of critical coarsening, which is the process by which dynamics progressively establishes the critical state~\cite{zeta2,cg2003}.

The present work is a sequel of \cite{cg2003} and of the related works \cite{zeta1,zeta2}.
Following the same line of thought, we investigate the dynamics of a different class of ZRP, for which the critical single-site distribution
at stationarity decays faster than a power law.
Again, analytical results can be obtained on the complete graph and in the thermodynamical limit.
The novelty of this case comes from the fact that, though
the asymptotic scaling functions associated to the single-site distributions are simpler than in the power-law case, the approach
to scaling is more complicated.
The analysis of this phenomenon is the main goal of the present work.

We proceed as in~\cite{cg2003},
analysing the coarsening dynamics of the system, first at criticality, then in the condensed phase. 
For both cases, we give an analytical treatment of the equations describing the temporal evolution of the single-site occupation probability
in the continuum scaling limit. 
For the critical phase, finite-time corrections to scaling can be explicitly determined.
The parallel study of the condensed phase turns out to be much harder. 
We use a semi-classical analysis of the differential 
equation describing the coarsening regime, in order to study the corrections
to scaling.

The same model was recently investigated in~\cite{gross}, with focus on coarsening in the condensed phase.
We shall review this work in the discussion at the end of the present paper.
A list of mathematical references on related issues can be found in~\cite{gross}.

\section{Definition of the model}

Consider a finite connected graph, made of $L$ sites, $i=1,\ldots,L$.
At time $t$, on each site we have $N_{i}(t)$ indistinguishable particles such that
\beq
\sum_{i=1}^{L}N_{i}(t)=N.
\eeq
The dynamics of the system is given by the rate $W(d,a,k,\l)$ at which a particle leaves
the departure site with label $d$, containing $N_{d}=k$ particles, and is transferred to the
arrival site with label $a$ containing $N_{a}=\l$ particles.
By definition of a ZRP, this hopping rate does not depend on the occupation of the arrival site and takes the simple form
\beq
W(d,a,k)=w_{d,a} u_k ,
\eeq
where $w_{d,a}$ accounts for diffusion from site $d$ to site $a$ and $u_k$ only depends on the occupation of the departure site.
In the present work we consider the complete graph where, by definition, all sites are connected.
We take $w_{d,a}=1/L$, i.e., all sites are equivalent and the system is spatially homogeneous.
We choose the rate
\beq\label{eq:uksig}
u_k=1+\frac{b}{k^\s},
\eeq
where $0<\s<1$ is an arbitrary exponent.
This form of the rate appeared in the past in several publications, such as~\cite{hanney,wis1,gl2005}.
It satisfies the criterion for condensation to occur~\cite{braz} for any value of $b$.
In contrast, for the ZRP with rate $u_k=1+b/k$, hereafter referred to as the $\s=1$ case, condensation only occurs for $b>2$.

We denote a configuration of the system at time $t$ by $\{n_i\}\equiv (n_1,n_2,\ldots,n_L)$,
where the $n_i=0,1,2,\dots$ are the values taken by the occupation numbers $N_{i}(t)$.
Thus, the complete knowledge of its dynamics involves the
determination of the probability $\prob(\{n_i\})$
of finding the system in the given configuration $\{n_i\}$ at time $t$. 
Hereafter we will focus our attention on a marginal of this distribution, namely on the probability of finding $k$ particles on
the generic site $i=1$, that, for short, we shall name the (single-site) occupation probability,
\beqa
f_{k}(t)=\prob(N_{1}(t)=k)=\langle\delta(N_1(t),k)\rangle
\\
=\sum_{n_1,\dots,n_{L}}\delta(n_1,k)\prob(\{n_i\}).
\eeqa
Conservation of probability and of density implies
\beqa
\label{eq:sumrule1}
\sum_{k\ge0}f_{k}(t) =1, \\
\sum_{k\ge1}k\,f_{k}(t) =\langle N_1(t)\rangle=\frac{N}{L}.
\eeqa
In the thermodynamic limit ($N\rightarrow \infty,
L\rightarrow \infty $, with fixed density $\rho =N/L$)
the last equation yields
\beq
\label{eq:sumrule2}
\sum_{k\ge1}k\,f_{k}(t) =\rho.
\eeq

\begin{figure}[htb]
\begin{center}
\includegraphics[angle=0,width=1\linewidth]{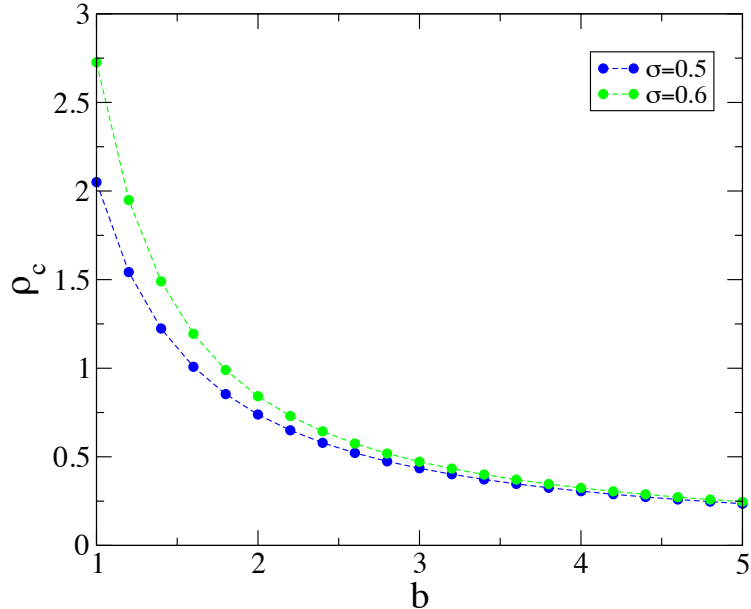}
\caption{\small
Critical density $\rho_c$ for the ZRP~(\ref{eq:uksig}) for two values of $\sigma$.
}
\label{fig:roc}
\end{center}
\end{figure}

Let us remind some well-known results on the stationary state of the ZRP~\cite{hanney,lux,gss}.
The weight of a configuration is given by
\beq
\prob(\{n_i\})=\frac{1}{Z_{L,N}}\prod_{i=1}^L p_{n_i},
\label{eq:pn}
\eeq
where
\beq
p_0=1,\qquad p_k=\frac{1}{u_1\dots u_k},
\label{eq:pk}
\eeq
and where the normalisation factor $Z_{L,N}$
reads
\beq
Z_{L,N}=\sum_{\{n_i\}}\prod_{i=1}^L p_{n_i}\,
\delta\left(\sum_{i=1}^L n_i,N\right).
\eeq
For the rate~(\ref{eq:uksig}) equation~(\ref{eq:pk}) leads to the estimate, for $k\gg1$,
\beq\label{eq:stretch}
p_k\sim\exp\left(-b\sum_{\ell=1}^k\frac{1}{\ell^\s}\right)
\sim\exp\left(-\frac{b}{1-\s}\,k^{1-\s}\right).
\label{pkas}
\eeq
In the thermodynamic limit, this ZRP is condensing whenever the density is larger than the critical density (see figure~\ref{fig:roc}),
\beq\label{eq:roc}
\rho_c=\frac{\sum_{k\ge1} k\,p_k}{\sum_{k\ge0} p_k}.
\eeq
The excess density $\rho-\rho_c$ corresponds to the condensate.
At the critical density, the occupation probabilities
\beq\label{eq:feqcrit}
f_{k,\eq}=\frac{p_k}{\sum_{k\ge0} p_k}
\eeq
decay as the stretched exponential law~(\ref{eq:stretch}).
Simple derivations of the above results can be obtained in the framework of the next section.

\section{Master equation}

From now on we consider the thermodynamic limit of the system on the complete graph.
In this mean-field geometry the temporal evolution of the occupation
probabilities $f_{k}(t)$ is explicitly given by the master equation
\beqa\label{eq:master}
\frac{ {\rm d} f_{k}(t)}{ {\rm d} t} &=&u_{k+1}\,f_{k+1}+\u\,f_{k-1}-
(u_{k}+\u)f_{k}\qquad (k\geq 1), \\
\label{eq:master0}
\frac{ {\rm d} f_{0}(t)}{ {\rm d} t} &=&u_{1}\,f_{1}-\u f_{0},
\eeqa
where
\beq\label{eq:ubart}
\u=\sum_{k=1}^{\infty }u_{k}f_{k}(t)
\eeq
is the rate at which a particle arrives on site number 1 from any other site.
It is the equation for a biased random walk for $N_{1}$ (or birth and death process)
on the positive integers $k=0,1,\ldots $. 
The rates of a jump to the right ($N_{1}=k\rightarrow N_{1}=k+1 $) or to the left ($N_{1}=k\rightarrow N_{1}=k-1 $) are respectively given by $\u$ and by $u_k$.
The equation for $f_{0}(t)$ is
special because one cannot select an empty site as a departure site, nor can 
$N_{1}$ be negative.
This random walk has the peculiar property of being constrained to have its average position fixed at the value $\rho$ (see~(\ref{eq:sumrule2})).

The form of the master equation~(\ref{eq:master}) is a direct consequence of the mean-field geometry.
It has the structure of the master equation for two sites~\cite{lux}, where the role of the second site is here played by the ensemble of all sites, through the self-consistency condition~(\ref{eq:ubart}).
This condition implies that (\ref{eq:master}) is non linear because the rate $\bar{u}_{t}$
is itself a function of the $f_{k}(t)$. 
Hence there is no
explicit solution of the master equation in closed form.
Yet one can extract
from~(\ref{eq:master}) an analytical description of the dynamics at long times, both at criticality and in the condensed phase, as will be seen in the next sections.
This master equation, (or closely related equations), appeared in previous works~\cite{zeta1,zeta2,cg2003} (see also~\cite{urn}).

In the stationary state we have
\beq\label{eq:defz}
\lim_{t\to\infty}\u=\sum_{k\ge1}u_k f_{k,\eq}=z,
\eeq
introducing the short notation $z$ for the limit.
Setting the left side of~(\ref{eq:master}) and (\ref{eq:master0}) to zero ($\dot{f}_{k}=0$) we obtain

\beq\label{eq:db}
\frac{f_{k+1,\eq}}{f_{k,\eq}}=\frac{z}{u_{k+1}},
\eeq
which expresses the detailed balance condition at equilibrium.
So
\beq
f_{k,\eq}=z^k p_k f_{0,\eq},
\eeq
where the $p_k$ are given by~(\ref{eq:pk}) and $f_{0,\eq}$ is fixed by the normalisation~(\ref{eq:sumrule1}).
Hence finally,
\beq
f_{k,\eq}=\frac{z^k p_k}{\sum_{k\ge0} z^k p_k}.
\eeq
Let $P(z)$ denote the generating series of the $p_k$ appearing in the denominator.
The second sum rule~(\ref{eq:sumrule2}) imposes that
\beq\label{eq:sumrule2+}
\rho=\frac{\sum_{k\ge1}z^k kp_k}{\sum_{k\ge0}z^k p_k}=\frac{zP'(z)}{P(z)}.
\eeq
This equation determines $z$ as a function of the density.
With the choice~(\ref{eq:uksig}), the $p_k$ decay as~(\ref{eq:stretch}), implying that the maximal value of the right side of~(\ref{eq:sumrule2+}), reached at $z=1$, is finite.
This finite value is the critical density~(\ref{eq:roc}),
\beq
\sum_{k\ge1}k f_{k,\eq}=\rho_c.
\eeq
We thus recover well-known results of the statics of the ZRP, either in the canonical or in the grand canonical formalisms, which are equivalent in the thermodynamical limit~\cite{hanney,gss}. 

Here we shall always keep time finite, even if very large, meaning that we are investigating the non-stationary regime, where the system stays homogeneous (in average, not configuration by configuration), allowing to follow the establishement of critical order, or to follow precursor effects of condensation.
Let us emphasize that~(\ref{eq:master}) can only account for the situation where all sites play the same role.
In other words, in the presence of a condensation transition, i.e., when $\rho>\rho_c$, this equation only accounts for the regime of the formation of the condensate.
Otherwise stated, in the thermodynamical limit, the non-stationary regime never ends, since the scale of time beyond which the stationary regime begins diverges with the system size $L$~\cite{cg2003,gl2005}.

\begin{figure}[htb]
\begin{center}
\includegraphics[angle=0,width=1\linewidth]{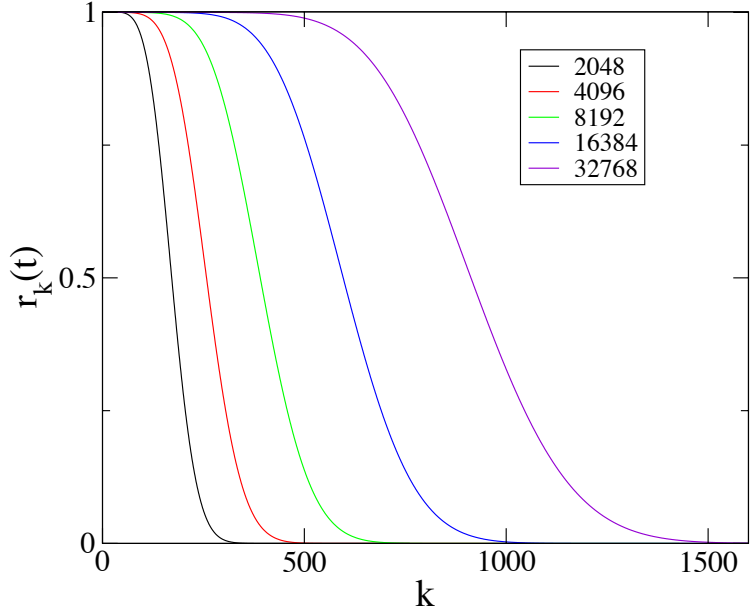}
\caption{\small
Critical ratio $r_k(t)=f_k(t)/f_{k,\eq}$ for $t=2048,4096,\dots,32768$, 
obtained by numerical integration of the discrete master equation (\ref{eq:master}).
Here $\sigma=0.6$, $b=1$.
}
\label{fig:rk}
\end{center}
\end{figure}
\begin{figure}[htb]
\begin{center}
\includegraphics[angle=0,width=1\linewidth]{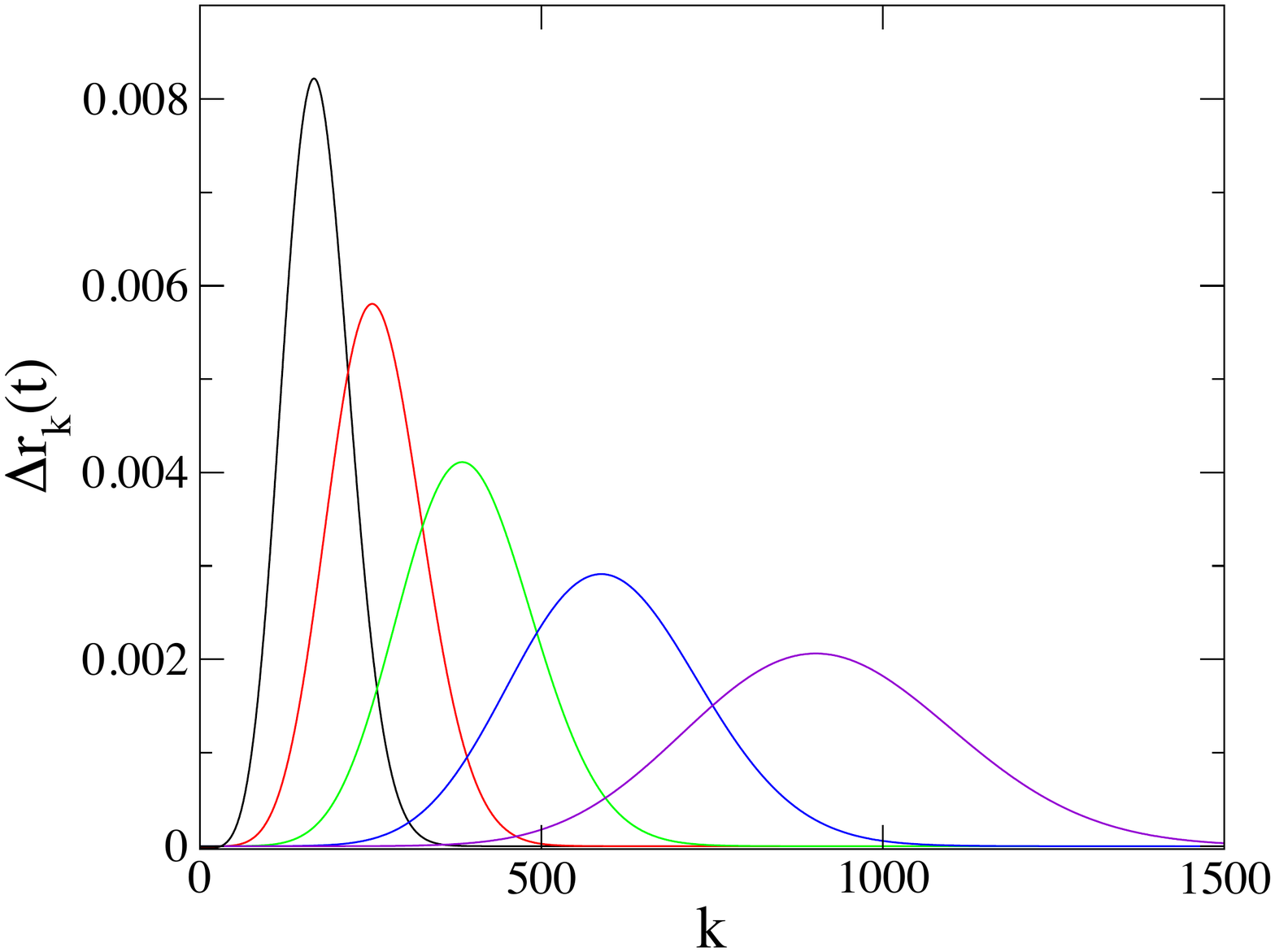}
\caption{\small
Differences $r_k(t)-r_{k+1}(t)$ for $t=2048,4096,\dots,32768$, 
obtained from the data of figure~\ref{fig:rk}.
}
\label{fig:Drk}
\end{center}
\end{figure}

%
\section{Critical coarsening}

We consider the ZRP with hopping rate $u_{k}$~(\ref{eq:uksig}), evolving on the complete graph from an homogeneous disordered initial condition specified by $f_k(0)$.
For instance, initially particles are distributed at random amongst sites, with an initial density $\rho =\rho_{c}$,
i.e., we consider a system with a Poissonian initial distribution of occupation probabilities,
\beq\label{eq:poiss}
f_k(0)=\e^{-\rho}\frac{\rho^k}{k!}.
\eeq
We investigate the critical coarsening process, i.e., the process by which dynamics progressively establishes the critical state.
The line of reasoning is similar to that followed in~\cite{zeta2,cg2003}.

Since the average rate $\bar{u}_{t}$ at which a particle leaves a generic site reaches its equilibrium value 
$z=1$ at large times, we set
\beq\label{eq:ubar-t}
\u= 1+\eta_{t},
\eeq
where the small scale $\eta_{t}$ will be determined hereafter.
Morevover, we are led to investigate the dynamics according to two time-occupancy regimes, as in~\cite{zeta2,cg2003}.
These regimes are defined as follows.

\subsubsection*{{\rm (I)} $k$ fixed, $t$ large} 

In this situation there is convergence to the equilibrium fluid phase.
Hence we set
\beq\label{eq:fkI_crit}
f_{k}(t)= f_{k,\eq}(1+w_{k}(t)), 
\eeq
with $f_{k,\eq}$ given by (\ref{eq:feqcrit}) and where the $w_{k}(t)$ are
proportional to $\eta_t$ as demonstrated below.

\subsubsection*{{\rm (II)} $k$ and $t$ are simultaneously large}

This is a regime where scaling is expected,
so we look for a solution to (\ref{eq:master}) of the
form
\beq\label{eq:fkII_crit}
f_{k}(t)= f_{k,\eq}\,\g(x,t),\qquad x=k\,\eps,
\eeq
where $\eps$ is a small scale, to be determined, $x$ is the scaling variable, and $\g(x,t)$ is expected to converge to the scaling function $g(x)$ in the limit of large times.

The present situation is closely related to critical coarsening for a ferromagnetic spin system quenched from infinite temperature down to $T_c$~\cite{bray,glcrit,revue}.
In such circumstances, spatial correlations develop in the system, just as in the critical state,
but only over a length scale which grows like $t^{1/z_c}$, where $z_c$ is the dynamic critical exponent.
On scales smaller than $t^{1/z_c}$ the system appears critical, while on larger scales the system is
still disordered. 
For instance, for Ising spins, $\s=\pm1$, the equal-time correlation function $C(r,t)=\langle\sigma_0(t)\sigma_{r}(t)\rangle$ scales
as
\beq
C(r,t) \approx r^{-2\beta/\nu} g\left(\frac{r}{t^{1/z_c}}\right),
\eeq
where $\beta$ and $\nu$ are the usual static exponents. 
The scaling function $g(x)$ goes to a constant as $x\to0$, while it falls off very
rapidly when $x\to\infty$.

Here, anticipating on what follows, starting from a homogeneous disordered initial condition, for a large but finite time~$t$,
and for $k$ much smaller than an ordering size of order $t^{1/z}$ (where the exponent $z$ is determined hereafter),
the system looks critical, i.e.,
the distribution $f_k(t)$ has essentially converged toward
the equilibrium distribution $f_{k,\eq}$.
To the contrary, for $k\gg t^{1/z}$, the system still looks disordered,
i.e., the $f_k(t)$ fall off very fast.
This is illustrated by figure~\ref{fig:rk}, obtained by numerical integration of~(\ref{eq:master}), which
depicts the ratio 
\beq\label{eq:rkt}
r_k(t)=\frac{f_k(t)}{f_{k,\eq}}.
\eeq
As time increases, this ratio exhibits a plateau of increasing length, reflecting the fact that the system equilibrates.
Then this ratio falls down very fast.

The agenda is now to determine the quantities $\eta_t$, $w_k(t)$, $\eps$, $g(x,t)$, using the sum rules~(\ref{eq:sumrule1}) and~(\ref{eq:sumrule2}), the master equations~(\ref{eq:master}) and (\ref{eq:master0}), 
and finally the assumptions~(\ref{eq:fkI_crit}) and (\ref{eq:fkII_crit}).
We proceed as follows.

\subsection{Time-occupancy regime (I)}

The expression~(\ref{eq:fkI_crit}) carried into~(\ref{eq:master}), (\ref{eq:master0}) imposes the left side $\dot{f}_{k}$ to vanish because it is proportional to $\dot{w_k}$, which is negligible compared to the right side of~(\ref{eq:master}), (\ref{eq:master0}). 
We thus obtain the quasi-stationary condition
\beq
\frac{f_{k+1,\eq}}{f_{k,\eq}}\frac{1+w_{k+1}}
{1+w_{k}}=\frac{1+\eta_t}{u_{k+1}},
\eeq
which formally resembles the detailed balance condition~(\ref{eq:db})
and yields $w_{k+1}-w_k=\eta_t$. 
Setting $w_k=v_k\eta_t$
we obtain 
\beq\label{eq:vk}
v_{k}=v_{0}+k,
\eeq
where $v_{0}$ is determined below (see~(\ref{eq:v0})).

\subsection{Time-occupancy regime (II)}

We now turn to the differential equation obeyed by $\g(x,t)$.
The ratio $r_k(t)=f_k(t)/f_{k,\eq}$ satisfies the equations
\beq\label{eq:rk}
\frac{ {\rm d} r_{k}(t)}{ {\rm d} t}=r_{k+1}+\u\,u_k\,r_{k-1}-
(\u+u_k)r_k \qquad(k\le 1),
\eeq
\beq
\frac{ {\rm d} r_{0}(t)}{ {\rm d} t}=r_{1}-\u r_0.
\eeq
Using~(\ref{eq:fkII_crit}), the left side of~(\ref{eq:rk}) becomes
\beq\label{eq:left}
\frac{ {\rm d} \g(x,t)}{ {\rm d} t}=\g'\frac{ {\rm d} x}{ {\rm d} t}+\dot\g=\g'k\dot\eps+\dot\g,
\eeq
and the right side yields
\beqa\label{eq:right}
 r_{k+1}+r_{k-1}-2r_k-\left(\eta_t+\frac{b}{k^{\s}}\right)(r_k-r_{k-1})+\eta_t\frac{b}{k^{\s}}r_{k-1}
\nonumber\\
\approx\eps^2\,\g'' -\left(\frac{\eta_t}{\eps^{\s}}+\frac{b}{x^{\s}}\right)\eps^{1+\s}\g' +\eta_t\eps^{\s}\frac{b}{x^{\s}}(\g+\cdots).
\eeqa
As will be shown below, the small scale $\eta_t$ is decaying exponentially fast.
Dropping the corresponding terms in the equation,
we obtain the partial differential equation
\beqa
\dot \g=
\eps^{2}\,\g''-\left(\frac{b}{x^{\s}}\eps^{1+\s}+x\frac{\dot\eps}{\eps}\right)\g'.
\eeqa
In order to equate the powers of $\eps$ in the second term of the right side, we set, (see also~(\ref{eq:epsilont})),
\beq\label{eq:epsilont1}
\eps= t^{-1/(1+\sigma)}.
\eeq 
We finally obtain the equation obeyed by $g(x,t)$,
\beq\label{eq:gc}
t\dot \g=t^{-a}\,\g''+\left(\frac{x}{1+\sigma}
-\frac{b}{x^\sigma}
\right)\g',
\eeq
with 
\beq\label{eq:defa}
a=\frac{1-\sigma}{1+\sigma}.
\eeq
By setting $\sigma=1$, we recover the equation found in~\cite{zeta2,cg2003} for the case $u_k=1+b/k$ (see~(\ref{eq:gcproto})), up to the left side of~(\ref{eq:gc}), which was omitted in these references, since the finite-time corrections need not be considered.
We now analyse equation~(\ref{eq:gc}).

\subsubsection*{(a) Scaling function}
At large times, the asymptotic scaling function
\beq
\g(x)=\lim\limits_{t \to \infty}\g(x,t)
\eeq
satisfies the equation
\beq
\left(\frac{x}{1+\sigma}-\frac{b}{x^\sigma}\right) \g'(x)
=0.
\eeq
Hence $\g(x)=1$ as long as the factor in parenthesis does not vanish.
This occurs for 
\beq\label{eq:x0}
x=x_0=[b(1+\sigma)]^{1/(1+\sigma)}.
\eeq
For $x>x_0$, $\g(x)=0$. 
The limiting scaling function is thus
a discontinuous curve, depicted in figure~\ref{fig:gx}.
In contrast, as can be seen on figure~\ref{fig:gx}, at finite time the solution of 
(\ref{eq:gc}) is a smooth curve, that we now investigate.

\subsubsection*{Remark}
Setting $\sigma=1$ in~(\ref{eq:x0}) yields $x_0=\sqrt{2b}$.
This prediction matches the result obtained for
the ZRP with rate $u_k=1+b/k$ in the limit $b\to\infty$~\cite{zeta2}.
Indeed, in this limit, the scaling function $g(x)$ becomes a discontinuous front at position $\sqrt{2b}$, as can be seen on its explicit expression~(\ref{eq:gc_anal}).\\
This matching between the two ZRP (corresponding respectively to the rates $u_k=1+b/k^\s$ and $u_k=1+b/k$) can be informally summarized as
\beq
\lim_{\s\to1}{\rm ZRP}_{\s<1}
\sim
\lim_{b\to\infty}{\rm ZRP}_{\s=1}.
\eeq
This property can be intuitively understood by noting that, in the limit $b\to\infty$, the decay of the $p_k\sim k^{-b}$ of the second ZRP (with $\s=1$) is formally faster than a power law,
thus falling into the class of the first ZRP (with $\s<1$).
The same matching will be encountered when investigating coarsening in the condensed phase 
(see the comment below~(\ref{eq:predict-w})).

\subsubsection*{(b) Finite-time corrections}

Consider, for a while, equation~(\ref{eq:gc}) without its left side.
This yields immediately
\beq\label{eq:psi}
-\g'(x,t)\propto\e^{-t^{a}\psi(x)},\quad\psi(x)=\int{\rm d} x\, \left(\frac{x}{1+\sigma}-\frac{b}{x^\sigma}
\right).
\eeq
It turns out that this expression does not account faithfully for the solution of the master equation~(\ref{eq:master}) as is demonstrated by what follows.
In other words, in order to account for the correct finite-time corrections to the scaling function $g(x)$, both terms $t\dot g$ and $t^{-a}g''$ should be kept.
However~(\ref{eq:psi}) puts us on the path of the correct ansatz for the solution
of~(\ref{eq:gc}).
Let us set
\beq\label{eq:phi}
-\g'(x,t)\propto\e^{-t^{a}\varphi(x)},
\eeq
where $\varphi(x_0)=0$, since $\g'(x,t)$ peaks at $x_0$ when $t\to\infty$.
If we substitute~(\ref{eq:phi}) in~(\ref{eq:gc})
we find a differential equation for the function $\varphi(x)$,
\beq\label{eq:Dphi}
\varphi'+a\frac{\varphi}{\varphi'}=\frac{x}{1+\sigma}-\frac{b}{x^\sigma}.
\eeq
This differential equation does not seem to be of a known type~\cite{zwil}.
Nevertheless the behaviours of $\varphi(x)$ at $x\to0$ or $x\to\infty$ can be predicted,
\beq
\varphi(x)\comport{\approx}{x\to0}{}\varphi(0)-\frac{bx^{1-\s}}{1-\s},\quad
\varphi(x)\comport{\approx}{x\to\infty}{}\frac{x^2}{4}.
\eeq
A plot of $\varphi(x)$, obtained by a numerical integration of~(\ref{eq:Dphi}), is given in figure~\ref{fig:phi}, which is in complete agreement with the data coming from the numerical integration of~(\ref{eq:master}), as will be commented later.

We can perform a local analysis of the behaviour of $\g'(x,t)$ around $x_0$, using the expansion
\beq
\varphi(x)\approx\frac{1}{2}\varphi''(x_0)(x-x_0)^2.
\eeq
Cast into~(\ref{eq:Dphi}), we obtain the relation
\beq
\varphi''(x_0)=1-\frac{a}{2}.
\eeq
Since, at large times, $\g(0,t)\to1$, we impose the normalisation
\beq
\int_0^\infty {\rm d} x\, \g'(x,t)=-1.
\eeq
So (\ref{eq:phi}) yields
\beq\label{eq:gprim2}
\g'(x,t)\approx-\sqrt{\frac{(1-a/2) t^{a}}{2\pi}}\e^{-(1-a/2)t^{a}(x-x_0)^2/2}.
\eeq
Hence
\beqa\label{eq:gxi}
\g(x,t)&\approx&-\int_{x}^{\infty} {\rm d} u\,\g'(u,t)=\frac{1}{2}\erfc \frac{\sqrt{(1-a/2)}\,t^{a/2}(x-x_0)}{\sqrt{2}}
\nonumber\\
&=&\frac{1}{2}\erfc \xi,
\eeqa
defining the new scaling variable $\xi$ as
\beq\label{eq:xi}
\xi=\frac{t^{a/2}(x-x_0)}{\sqrt{2/(1-a/2)}}
=\frac{k-k_{\typ}}{\sqrt{2t/(1-a/2)}},
\eeq
and where the typical location of the front depicted in figure~\ref{fig:rk} is
\beq
k_{\typ}=x_0 t^{1/(1+\sigma)}.
\eeq
So, the front moves more rapidly (as $t^{1/(1+\sigma)}$) 
than it widens (as $t^{1/2}$).
The difference between the two exponents is
\beq
\frac{1}{1+\sigma}-\frac{1}{2}=\frac{a}{2},
\eeq
which is precisely the exponent appearing in the scaling variable $\xi$.
To summarize, the bulk of the function $\g(x,t)$ is given by $(1/2)\erfc \xi$, hence $-\g'(x,t)$ is a Gaussian, and the large deviations of the latter are given by~(\ref{eq:phi}), where $\varphi(x)$ is the large-deviation function.

\subsubsection*{Remark}
For $k$ and $t$ large, using the results above, we get
$f_k(t)\sim \e^{-k^2/(4t)}$,
as for the case $u_k=1+b/k$, see Appendix.

\begin{figure}[htb]
\begin{center}
\includegraphics[angle=0,width=1\linewidth]{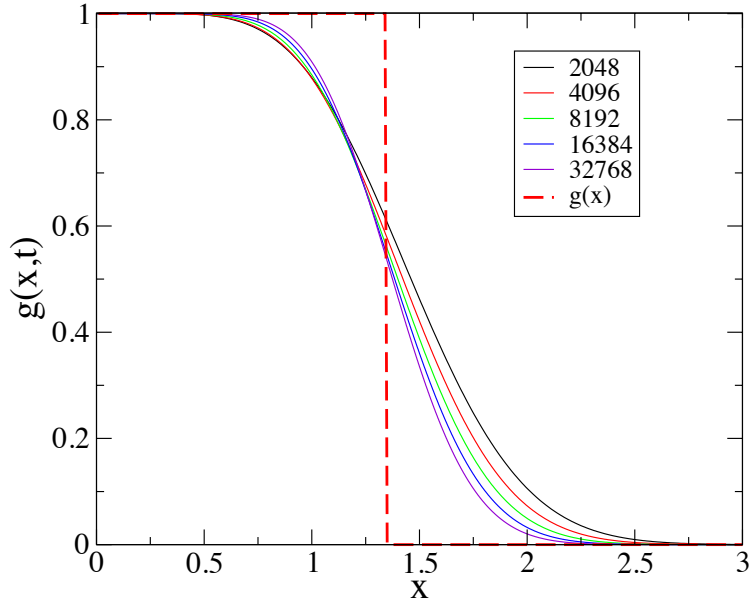}
\caption{\small
Function $\g(x,t)$ against $x$ for $t=2048,4096,\dots,32768$, 
obtained from the data of figure~\ref{fig:rk}.
The discontinuous curve is the limiting scaling function $\g(x)$. 
The discontinuity is at $x_0=(b(1+\sigma))^{1/(1+\sigma)}$, i.e., $x_0=1.341\dots$ with $\sigma=0.6$, $b=1$.
}
\label{fig:gx}
\end{center}
\end{figure}

\begin{figure}[htb]
\begin{center}
\includegraphics[angle=0,width=1\linewidth]{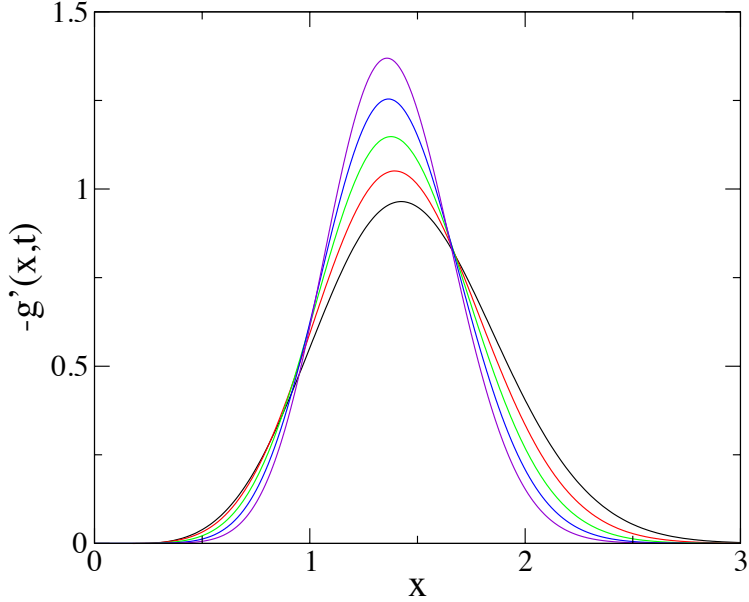}
\caption{\small
Function $-\g'(x,t)$ for $t=2048,4096,\dots,32768$, 
obtained from the data of figure~\ref{fig:gx}.
}
\label{fig:gprime}
\end{center}
\end{figure}

\begin{figure}[htb]
\begin{center}
\includegraphics[angle=0,width=1\linewidth]{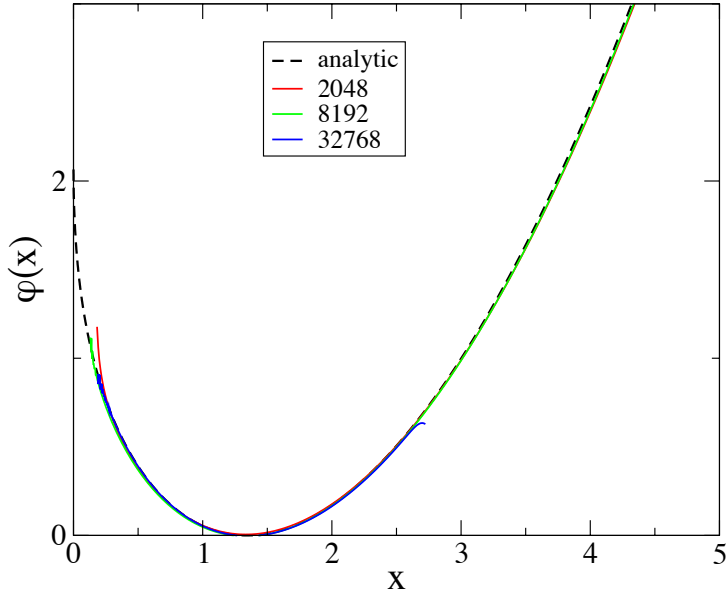}
\caption{\small
Comparison between the theoretical prediction for the large-deviation function $\varphi(x)$ given by~(\ref{eq:Dphi})
(dashed) with the results obtained from the data of figure~\ref{fig:gprime}, using the definition~(\ref{eq:phi}).
These data have been shifted to $x_0$.
}
\label{fig:phi}
\end{center}
\end{figure}

\begin{figure}[htb]
\begin{center}
\includegraphics[angle=0,width=1\linewidth]{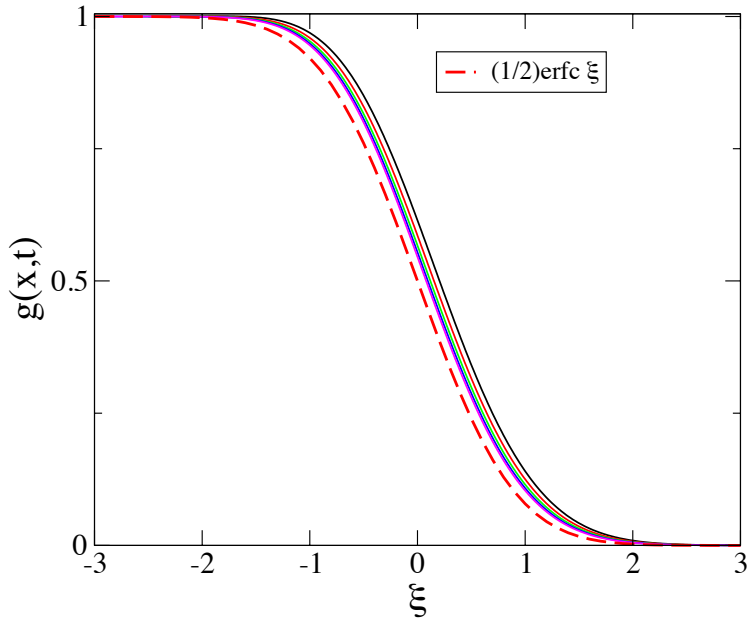}
\caption{\small
Same data as in figure~\ref{fig:rk}, against scaling variable $\xi$ defined in~(\ref{eq:xi}), compared to the bulk scaling function $(1/2)\erfc \xi$ (dashes).
}
\label{fig:xi06}
\end{center}
\end{figure}
\begin{figure}[htb]
\begin{center}
\includegraphics[angle=0,width=1\linewidth]{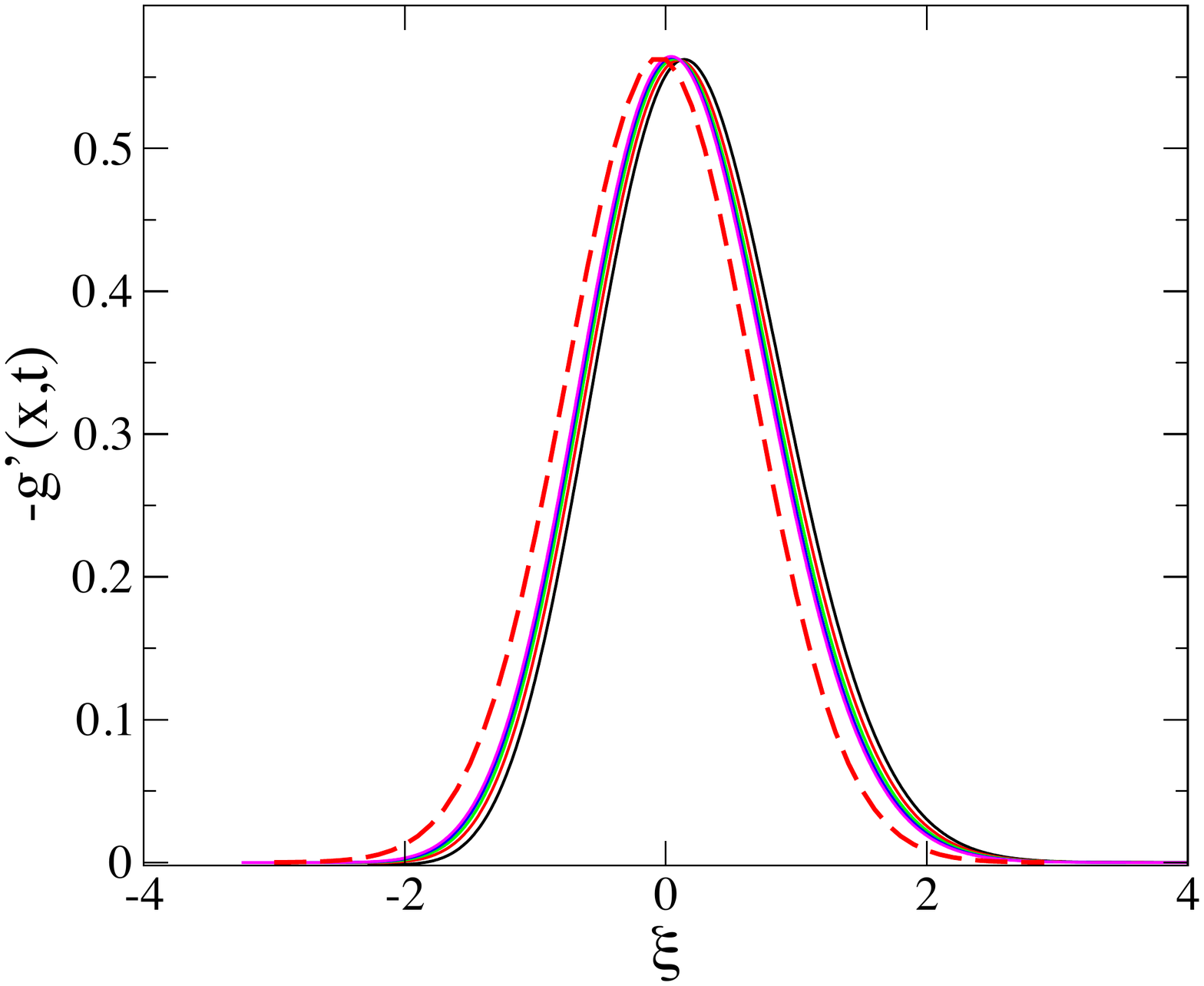}
\caption{\small
Same data as in figure~\ref{fig:Drk} against $\xi$, compared to the Gaussian $\e^{-\xi^2}/\sqrt{\pi}$ (dashes).
This function is normalised to unity with respect to $\xi$.
}
\label{fig:Dxi06}
\end{center}
\end{figure}

%
\subsection{Exact numerical results}

We now compare the theoretical predictions above to the results of numerical integrations
of the discrete master equation (\ref{eq:master}). 
\begin{enumerate}
\item Figure~\ref{fig:rk}, already commented upon above, depicts $r_k(t)$ for the five different times
$t=2048$, $4096$, $8192$, $16384$, $32768$, for $\sigma=0.6$, $b=1$.
\item Figure~\ref{fig:Drk} depicts the forward differences, $r_k(t)-r_{k+1}(t)$, of the previous data.
\item Figure~\ref{fig:gx} depicts the data of figure~\ref{fig:rk} plotted against $x$, i.e., $r_k(t)\approx\g(x,t)$.
The discontinuous curve is the limiting scaling function $\g(x)$. 
\item Figure~\ref{fig:gprime} depicts the derivative $-g'(x,t)$ 
obtained by plotting the data of figure~\ref{fig:Drk} against $x$.
The location of the maximum is moving towards $x_0=1.341...$.
The successive curves peak as $t^{a/2}$ towards the limiting function $\delta(x-x_0)$.
\item Figure~\ref{fig:phi} gives a comparison between the theoretical prediction for $\varphi(x)$, obtained by a numerical integration of~(\ref{eq:Dphi}),
with the results obtained from the data of figure~\ref{fig:gx}, using the definition~(\ref{eq:phi}).
Note the perfect collapse of the latter onto a master curve, as well as their adequation with the former.
\item Figure~\ref{fig:xi06} and~\ref{fig:Dxi06} depict $\g(x,t)$ and $-\g'(x,t)$ against $\xi$, respectively, demonstrating the convergence of the data towards the theoretical prediction~(\ref{eq:gxi}) and its derivative $\e^{-\xi^2}/\sqrt{\pi}$.

\end{enumerate}

\subsection{Using the sum rules}

Taking into account the respective contributions of the two time-occupancy regimes (I) and (II),
the sum rules (\ref{eq:sumrule1}) and (\ref{eq:sumrule2}) lead respectively to the following relationships,
\beq\label{eq:sr1}
0=\sum_{k\ge0}(f_k(t)-f_{k,\eq})\approx
\eta_t(v_0+\rho_c)-I_0,
\eeq
and
\beq\label{eq:sr2}
0=\sum_{k\ge1} k(f_k(t)-f_{k,\eq})\approx
\eta_t(v_0 \rho_c+\mu_c)-I_1,
\eeq
where $\mu_c=\langle N_1^2\rangle=\sum k^2 f_{k,\eq}$,
and where the integrals $I_0$ and $I_1$ read
\beqa\label{i0}
 I_0&=& \int_{0}^\infty {\rm d} k\,f_{k,\eq}
 (1-\g(k\eps,t)),\\
 I_1&=&
 \int_{0}^\infty {\rm d} k\,k\,f_{k,\eq}
 (1-\g(k\eps,t)).
\eeqa
The integral $I_0$ is negligible compared to $I_1$ since the integrand of the latter bears an additional factor $k$, which is large.
So, comparing 
(\ref{eq:sr1}) and (\ref{eq:sr2}), we conclude that
\beq\label{eq:v0}
v_{0}+\rho _{c} =0.
\eeq
So the proportionality constant between $\eta_t$ and $I_1$ in (\ref{eq:sr2}) is the variance $\mu_c-\rho_c^2$.

We can now address the question of the time dependence of $\eta_t$.
We have seen that in regime (I) (for $k$ fixed, $t\to \infty$) $r_{k+1}-r_k\approx \eta_t$.
On the other hand,  in regime (II), for $x\to0$, we have $r_{k+1}-r_k\sim \g'(0,t)$.
Since a matching mechanism between the two time-occupancy regimes (I) and (II) should take place, it is natural to suppose that
\beq\label{eq:timedep}
\eta_t\sim \e^{-t^{a}\,\varphi(0)}.
\eeq
This result has been checked numerically.
Looking at (\ref{eq:sr2}), we infer that
\beq
\eta_t\sim I_1\sim \e^{-t^{a}\,\varphi(0)}.
\eeq

\section{Coarsening dynamics in the condensed phase}
\label{sec:cond_mf}

In this section we describe the dynamics of the ZRP in the condensed phase.
As above, the system evolves from a homogeneous disordered initial condition, given e.g. by~(\ref{eq:poiss}), now with an initial density $\rho >\rho_{c}$.

\subsection{Time-occupancy regimes}
\begin{figure}[htb]
\begin{center}
\includegraphics[angle=0,width=1\linewidth]{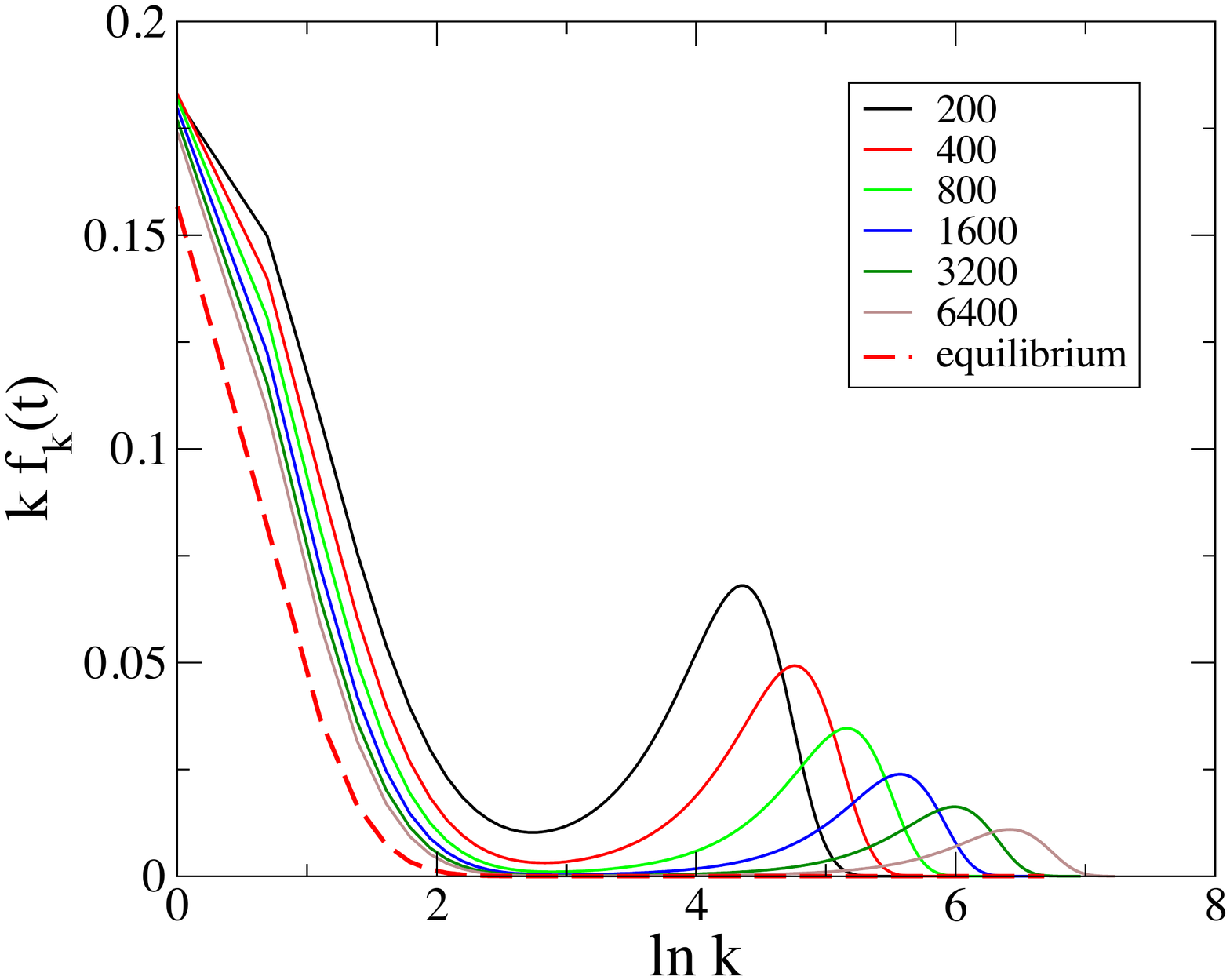}
\caption{\small 
Product $kf_k(t)$ against $\ln k$ for $t=200, 400, \dots, 6400$.
The dashed curve is the equilibrium distribution $f_{k,\eq}$.
This figure substantiates the separation between the two time-occupancy regimes (I) and (II).
Here $\s=1/2,b=4$, $\rho_c\approx0.306$, $\rho=20\rho_c$.
}
\label{fig:newfk}
\end{center}
\end{figure}

\begin{figure}[htb]
\begin{center}
\includegraphics[angle=0,width=1\linewidth]{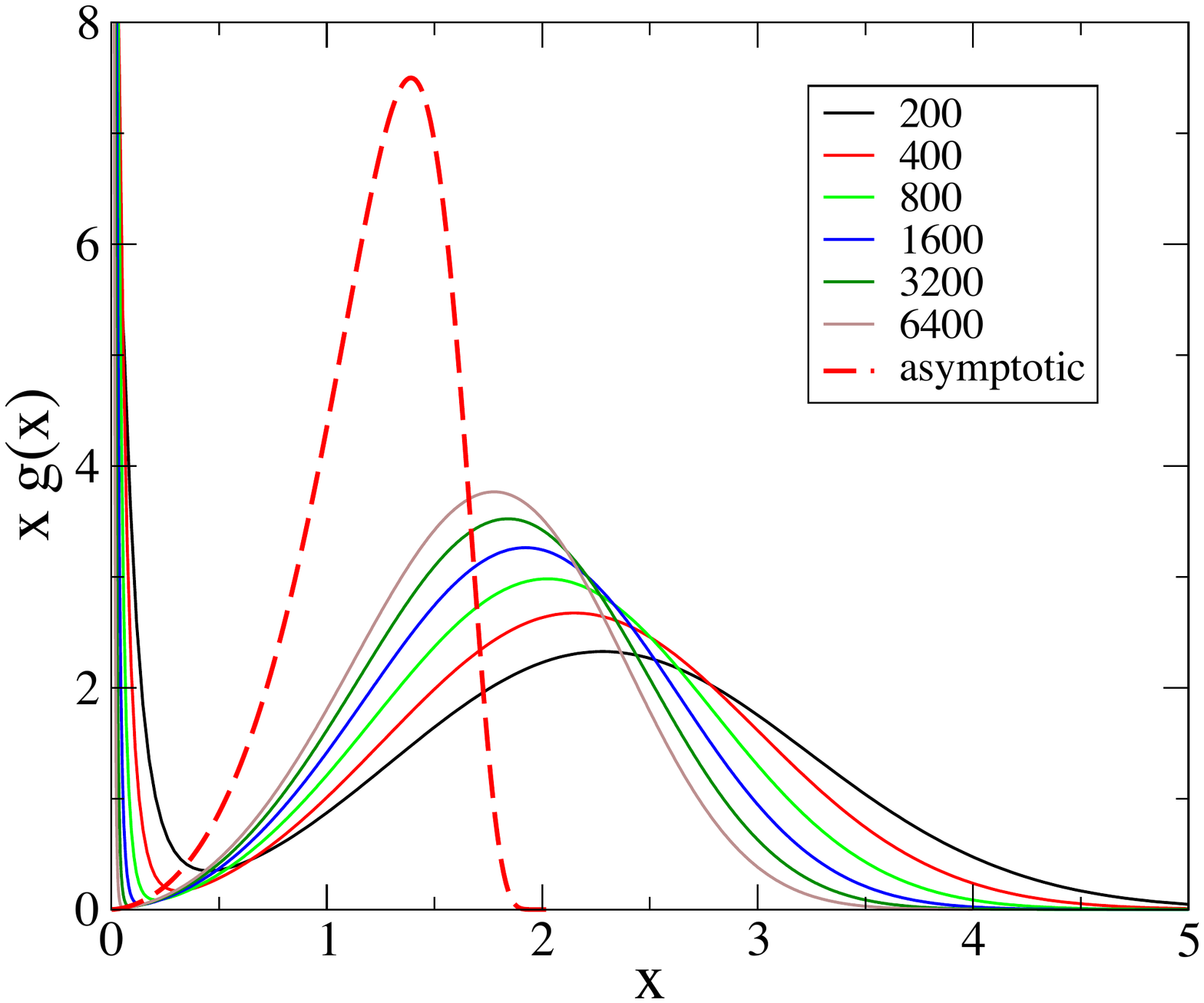}
\caption{\small Same as figure~\ref{fig:newfk} against the scaling variable $x$.
The asymptotic dashed curve $xg(x)$~(\ref{eq:gstat}) is normalised according to~(\ref{eq:rs2_cond}).
}
\label{fig:newxgx}
\end{center}
\end{figure}

Figure~\ref{fig:newfk}, obtained by numerical integration of the master equation~(\ref{eq:master}), depicts $kf_k(t)$ against $\ln k$ for increasing times.
These curves exhibit two well separated time-occupancy regimes:
a first regime of convergence to equilibrium, before the dip, then a second regime corresponding to the bumps shifting progressively to the right.
Rescaling $k$, as detailed below, we obtain figure~\ref{fig:newxgx} which indicates a slow convergence to a limiting curve (dashes).


These observations can be formalized as follows.
We still set
\beq\label{eq:ubar-t+}
\u= 1+\eta_{t},
\eeq
where the small scale $\eta_{t}$ will turn out to be different from its critical counterpart~(\ref{eq:ubar-t}),
and we investigate the dynamics according to two time-occupancy regimes, as for the critical case.

\subsubsection*{{\rm (I)} $k$ fixed, $t$ large} 

In this situation there is convergence to the equilibrium fluid phase and we still set
\beq\label{eq:fkI}
f_{k}(t)= f_{k,\eq}(1+w_{k}(t)).
\eeq
By the same reasoning as for the critical case we find, with $w_k=v_k\eta_t$, that
\beq\label{eq:vk+}
v_{k}=v_{0}+k,
\eeq
with $v_{0}$ given by~(\ref{eq:rs1_cond}), as demonstrated below.

\subsubsection*{{\rm (II)} $k$ and $t$ are simultaneously large}

In the spirit of~\cite{zeta2,cg2003} (see appendix), we look for a solution of (\ref{eq:master}) of the form
\beq\label{eq:fkscalcond}
f_{k}(t)= \eps^{2}g(x,t),\quad x=k\eps,
\eeq
where $x$ is the scaling variable and $\eps$ is a small scale which will turn out to be given again by~(\ref{eq:epsilont1}).
As in the critical case, $g(x,t)$ is expected to converge, in the limit of large times, to the scaling function $g(x)$, to be determined.
In~\cite{zeta2,cg2003} the explicit time dependence of $g(x,t)$ was not taken into account because the finite-time corrections were inessential.

\subsection{Using the sum rules}
We start by using~(\ref{eq:fkI}) and (\ref{eq:fkscalcond}) as well as the sum rules~(\ref{eq:sumrule1}) and~(\ref{eq:sumrule2}) 
in order to derive~(\ref{eq:rs1_cond}) and~(\ref{eq:rs2_cond}).
We proceed as follows.
Let us mark the separation between the two time-occupancy domains by the position of the dip $k_{\star}$ clearly seen on 
figure~\ref{fig:newfk}.
The first sum rule~(\ref{eq:sumrule1}) leads to 
\beq
1=\sum_{k=0}^{k_{\star}}f_{k,\eq}\left(1+\eta_t v_k\right)
+\eps^2\sum_{k=k_{\star}}^\infty g(k\eps,t),
\eeq
hence
\beq\label{eq:xx}
\eta_t
\sum_{k=0}^{k_{\star}}f_{k,\eq}(v_0+k)
=-\eps\int_{x_{\star}}^\infty {\rm d} x\,g(x,t)+\sum_{k=k_{\star}}^\infty f_{k,\eq},
\eeq
where $x_{\star}=k_{\star}\eps^{-1}$.
We then let $k_{\star}\to\infty$ and $x_{\star}\to0$.
This is justified by the fact that $k_{\star}$ increases much slower in time than $\eps^{-1}$, as a simple argument shows (see the remark below).
Thus
\beq\label{eq:rs1-0}
\eta_t(v_{0}+\rho_{c})=-\eps\int_{0}^\infty {\rm d} x\,g(x,t).
\eeq
The second sum rule~(\ref{eq:sumrule2}) yields
\beq
\rho=\sum_{k=0}^{k_{\star}}kf_{k,\eq}\left(1+\eta_t v_k\right)
+\eps^2\sum_{k=k_{\star}}^\infty kg(k\eps,t),
\eeq
hence
\beq
\rho-\sum_{k=0}^{k_{\star}}kf_{k,\eq}=
\eta_t
\sum_{k=0}^{k_{\star}}kf_{k,\eq}(v_0+k)
+\int_{x_{\star}}^\infty {\rm d} x\,xg(x,t)
\eeq
Thus
\beq\label{eq:rs2-0}
\rho -\rho_{c}=\eta_t (v_0\rho_c+\mu_c)+\int_{0}^{\infty } {\rm d} x\,xg(x,t). 
\eeq
Taken together and assuming, as shown below, that $\eta_t\gg\eps$, these equations impose the two constraints
\beq\label{eq:rs1_cond}
v_{0}+\rho_{c}=0,
\eeq
and 
\beq\label{eq:rs2_cond}
\rho -\rho_{c}=\int_{0}^{\infty } {\rm d} x\,xg(x,t). 
\eeq
Equation~(\ref{eq:rs1_cond}) determines $v_0$, while~(\ref{eq:rs2_cond}) gives the normalisation of the function $g(x,t)$ that we know investigate.

\subsubsection*{Remark}
In order to estimate the time dependence of $k_{\star}$ we impose the derivative of $f_k(t)$ with respect to $k$ to vanish at this point.
Using (\ref{eq:stretch}) and the fact that $g(x,t)\sim x^{\s}$ at large time (see section~\ref{sec:scalingfctn}), we obtain
\beq
\e^{-b\,k_{\star}^{1-\s}/(1-\s)}\sim \eps^{2+\s},
\eeq
hence
\beq
k_{\star}\sim(\ln t)^{1/(1-\s)},
\eeq
which is well verified numerically.

\subsection{Equation satisfied by $g(x,t)$}
Inserted into~(\ref{eq:master}) the scaling form~(\ref{eq:fkscalcond}) yields a linear differential equation for $g(x,t)$.
Indeed~(\ref{eq:master}) can be rewritten as
\beq
\fl\frac{ {\rm d} f_k(t)}{ {\rm d} t}
=f_{k+1}+f_{k-1}-2f_k+b\left(\frac{f_{k+1}}{(k+1)^{\sigma}}-\frac{f_{k}}{k^{\sigma}}\right)
-\eta_t(f_k-f_{k-1}).
\eeq
Replacing the discrete derivatives of $f_k$ with respect to $k$ by derivatives of $g(x,t)$ with respect to $x$, yields
\beq
\fl\dot{\eps}\eps(2g+xg')+\eps^2\dot g=
\eps^4g''+\eps^{3+\sigma}\left(\frac{b}{x^\sigma}-\frac{\eta_t}{\eps^\sigma}\right)g'-\eps^{3+\sigma}\frac{\s b}{x^{1+\sigma}}g.
\eeq
We divide both sides by $\eps^{3+\sigma}$.
Setting, as for the critical case,
\beq\label{eq:epsilont}
\eps= t^{-1/(1+\sigma)},
\eeq 
implies $\dot{\eps}/\eps^{2+\sigma}=-1/(1+\sigma)$.
Finally setting
\beq\label{eq:eta-eps}
\eta_t=A\,\eps^\sigma,
\eeq
where it is understood that $A$ is a function of $t$,
we obtain the continuum equation
\beq\label{eq:edpCond}
t\dot g=
t^{-a}g''+\left(\frac{x}{1+\sigma}-A+\frac{b}{x^\sigma}\right)g'+\left(\frac{2}{1+\sigma}-\frac{\s b}{x^{1+\sigma}}\right)g,
\eeq
with definition~(\ref{eq:defa}) for the exponent $a$.
For $\sigma=1$, omitting the left side of~(\ref{eq:edpCond}) we recover the equation corresponding to the ZRP with rate $u_k=1+b/k$, see~(\ref{app:gcond}).
For this latter case, the term $t\dot g$ would give a finite-time correction to the scaling function $g(x)$.
This was not considered in~\cite{zeta1,zeta2,cg2003} because the convergence to $g(x)$ was very fast.

\subsubsection*{Remark}
A heuristic argument confirming the relationship~(\ref{eq:eta-eps}) between $\eta_t$ and $\eps$ is as follows.
First, balancing $\u$ with $u_k$ yields
$k\sim \eta_t^{-1/\sigma}$.
Then, noting that $k\sim\eps^{-1}$ in the scaling region,~(\ref{eq:eta-eps}) ensues.

\section{Analysis of the continuum equation~(\ref{eq:edpCond})} 

\subsection{The scaling function}
\label{sec:scalingfctn}

The stationary solution $g(x)$ of~(\ref{eq:edpCond}), obtained by letting $t\to\infty$, satisfies the equation
\beq\label{eq:eq-diff-cond}
\left(\frac{x}{1+\sigma}-A+\frac{b}{x^\sigma}\right)g'+\left(\frac{2}{1+\sigma}-\frac{\s b}{x^{1+\sigma}}\right)g=0,
\eeq
which can be rewritten as
\beq
D(x) g' +\left(D'(x)+\frac{1}{1+\s}\right)g=0,
\eeq
where
\beq
D(x)=\frac{x}{1+\sigma}-A+\frac{b}{x^\sigma}.
\eeq
The solution of this equation is
\beq\label{eq:gstat}
g(x)\propto\frac{1}{D(x)}\exp\left(-\frac{1}{1+\s}\int_{0}^{x} {\rm d} u \frac{1}{D(u)}\right).
\eeq
If $x\to0$ then $D(x)\sim x^{-\s}$, hence $g(x)\sim x^{\s}$.
If $x\to\infty$ then $D(x)\sim x/(1+\s)$, thus
\beq
g(x)\sim \frac{1}{x}\e^{-\int^{x} {\rm d} u/u}\sim x^{-2},
\eeq
and therefore $\int {\rm d} x\,x g(x)$ diverges.
This therefore rules out the possibility for $g(x)$ to have a support extending to infinity.

Let us more generally discuss which value of the amplitude $A$ is selected.
Denoting the value of $x$ such that $D(x)$ is minimum by 
\beq
x_{0}=\left[b\,\s(1+\s)\right]^{\frac{1}{1+\s}},
\eeq
we have
\beq\label{eq:Dumini}
D(x_{0})=A_0-A,
\eeq
where
\beq
A_0=\frac{x_{0}}{\s}.
\eeq
Three cases are to be considered according to the sign of $D(x_{0})$:
\begin{enumerate}
\item $D(x_{0})>0$, or $A<A_0$,
\item $D(x_{0})=0$, or $A=A_0$,
\item $D(x_{0})<0$, or $A>A_0$. 
\end{enumerate}
Case (i) is necessarily ruled out since the support of $g(x)$ would extend to infinity.
In case (iii) $g(x)$ would vanish at $x_1$, the first zero of $D(x)$.
As will be demonstrated below, the selected solution is case (ii).
For the latter, the asymptotic scaling function $g(x)$ has an essential singularity at $x_{0}$,
\beq
g(x)\sim \e^{-2 x_0/((1+\s)(x_0-x))},
\eeq
and vanishes for $x>x_0$.
The integral~(\ref{eq:gstat}) is explicit when $\s$ is rational.
We plot $x g(x)$ for $\s=1/2$ and $b=4$ in figure~\ref{fig:newxgx}, with the normalisation~(\ref{eq:rs2_cond}).
Figure~\ref{fig:newxgx} also depicts $xg(x,t)$ for several values of time.
One observes slow convergence to the asymptotic function $xg(x)$, that we now analyse.

\subsection{A simplified form of~(\ref{eq:edpCond})}

Let us now analyse a simplified form of~(\ref{eq:edpCond}) where the left side, i.e., the $t \dot g$ term, is omitted.
We rewrite the equation for convenience,
\beq\label{eq:eqdif-first}
\gamma g''+\left(\frac{x}{1+\sigma}-A+\frac{b}{x^\sigma}\right)g'+\left(\frac{2}{1+\sigma}-\frac{\s b}{x^{1+\sigma}}\right)g=0,
\eeq
where we have set $\gamma=t^{-a}$.
Since the time dependence of $g(x,t)$ only enters through the small parameter $\gam$,
we denote the solution of~(\ref{eq:eqdif-first}) by $g(x,\gam)$.
The constraints on this function are that it should be positive and vanish at zero and infinity.
This selects an amplitude $A$ depending on $\gam$, with limiting value $A_0$, when $\gamma\to0$, as we now show.

We first cast~(\ref{eq:eqdif-first}) in normal form by setting $g(x,\gam)=v(x)w(x)$, with $v(x)$ chosen in such a way that 
the resulting equation for $w$ has no first derivative term.
This gives
\beq
v(x)=\exp\left(-\frac{1}{2\gam}\int_{0}^{x} {\rm d} u\,D(u)\right),
\eeq
with
\beq
\int_{0}^{x} {\rm d} u\,D(u)=x\left(\frac{x}{2(1+\s)}-A+\frac{b}{(1-\s)x^{\s}}\right).
\eeq
So
\beqa
v(x)\comport{\sim}{x\to\infty}{}\exp\left(-\frac{x^2}{4\gam(1+\s)}\right),
\nonumber\\
v(x)\comport{\sim}{x\to0}{}\exp\left(-\frac{b\, x^{1-\s}}{2\gam(1-\s)}\right).
\eeqa
The equation for $w(x)$ reads
\beq\label{eq:w}
\gam^2 w''
=
\left(\frac{D(x)^2}{4}-\gam\left(\frac{D'(x)}{2}+\frac{1}{1+\s}\right)\right)w,
\eeq
which is a Schr\"odinger equation 
\beq\label{eq:schrod}
\left(-\gam^2\frac{\d^2}{ {\rm d} x^2}+V(x,A)\right)w=0,
\eeq
with potential
\beq
V(x,A)=\frac{1}{4}\left(\frac{x}{1+\sigma}-A+\frac{b}{x^\sigma}\right)^2-\frac{\gam}{2}\left(\frac{3}{1+\sigma}-\frac{\s b}{x^{1+\sigma}}\right).
\eeq

\begin{figure}[htb]
\begin{center}
\includegraphics[angle=0,width=1\linewidth]{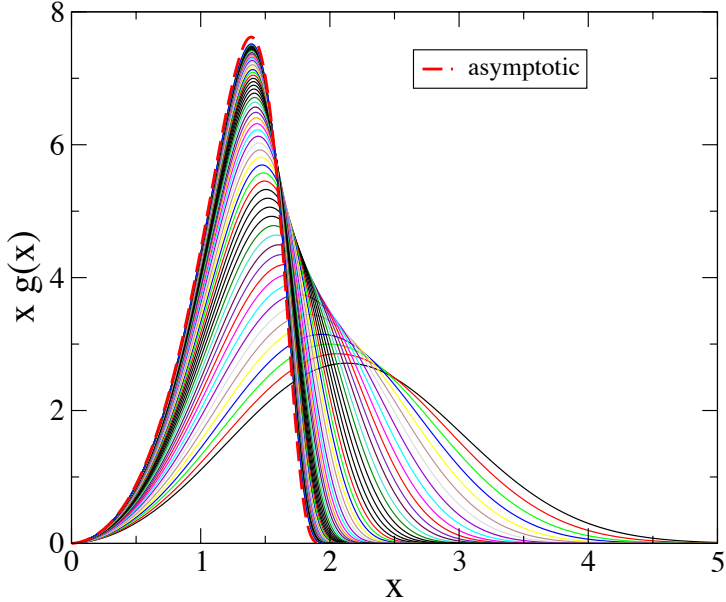}
\caption{\small Solutions of the differential equation~(\ref{eq:eqdif-first}) obtained by numerical integration, for decreasing values of $\gam$ ($\s=1/2, b=4$), ranging from $\gam=0.4$ ($t\approx 15$) to $\gam\approx 8.5\, 10^{-6}$ ($t\approx 1.62\,10^{15}$).
}
\label{fig:g+}
\end{center}
\end{figure}

\begin{figure}[htb]
\begin{center}
\includegraphics[angle=0,width=1\linewidth]{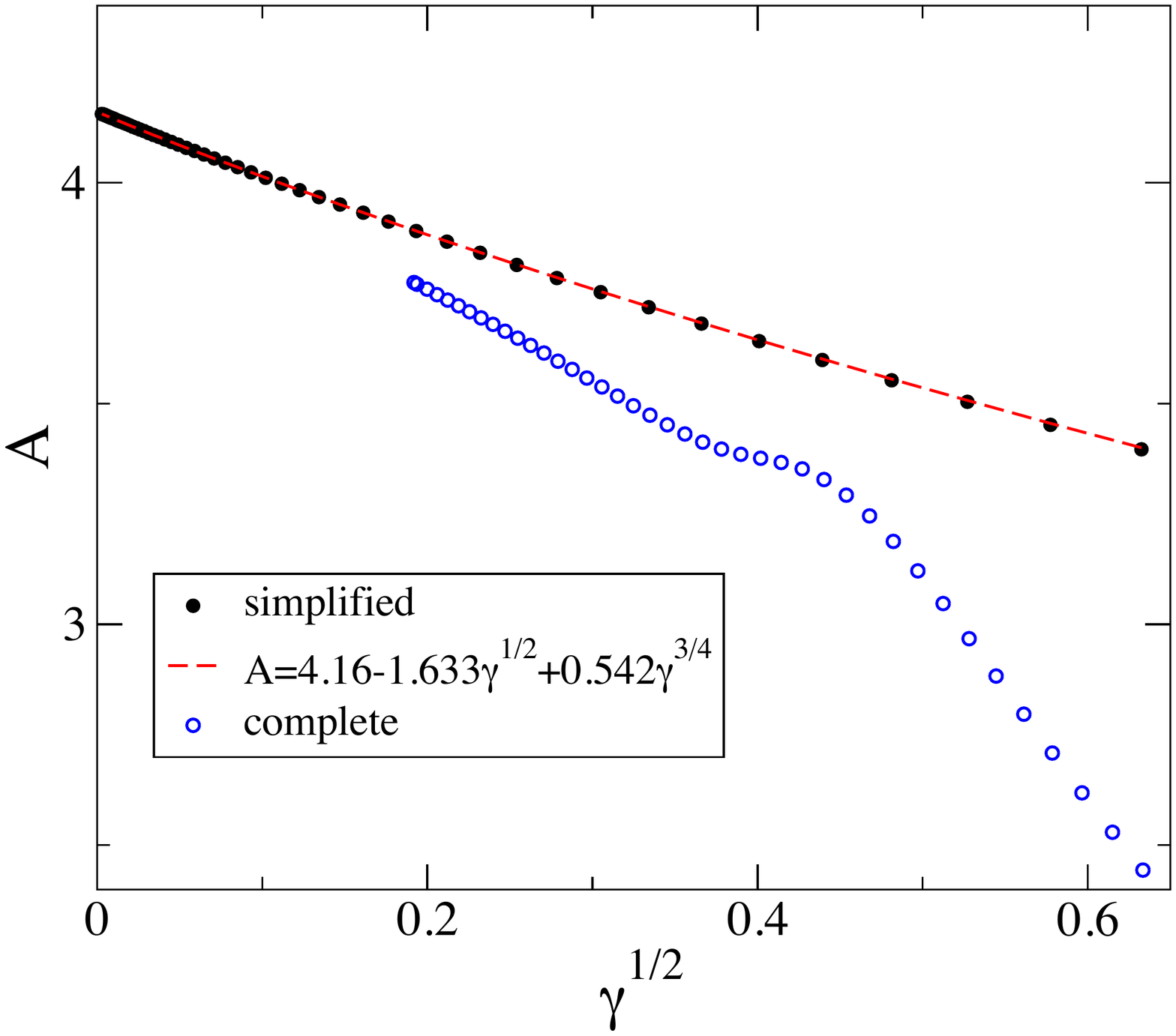}
\caption{\small Upper curve: comparison between the amplitude $A$ corresponding to the solutions, depicted in figure~\ref{fig:g+}, of the simplified equation~(\ref{eq:eqdif-first}) (black dots) and the prediction~(\ref{eq:Aprediction}) (red dashes).
Lower curve: amplitude $A$ obtained by numerical integration of the master equation~(\ref{eq:master}) (the continuum limit of which is the complete equation~(\ref{eq:edpCond})) (open blue dots).
}
\label{fig:Anew+}
\end{center}
\end{figure}

We now analyse this equation in the semi-classical regime $\gam\to0$.
In the ground state ($w(x)$ is positive), the particle sits at the minimum of the potential.
Moreover, since this solution has zero energy,
we have to express that this minimum vanishes, at leading order.

We expect this minimum to be located in the vicinity of $x_0$, with $A$ close to $A_0$
because for $\gam=0$, $V(x,A)$ is minimum at $x_0$ (see~(\ref{eq:Dumini})), and vanishes if $A=A_0$.
We have
\beq
V(x_0,A)=\frac{(A-A_0)^2}{4}-\gam\frac{1}{1+\s}.
\eeq
Hence choosing 
$A=A_1\equiv A_0-2\gam^{1/2}/\sqrt{1+\s}$ 
yields $V(x_0,A_1)=0$, and $V'(x_0,A_1)=-\gam/(2x_0)$ (this last value is independent of $A$).
We refine this analysis by expanding the potential around its minimum.
We set 
\beqa
x&=&x_0+\gam^{3/8}z,
\\
A&=&A_0-\alpha\gam^{1/2}+\beta\gam^{3/4},
\eeqa
where $\alpha$ and $\beta$ will be determined.
We obtain, for the potential, now a function of $z$,
\beq
\tilde V(z)=\frac{\gam}{4}\left(\alpha^2-\frac{4}{1+\s}\right)
+\gam^{5/4}\frac{\alpha}{4x_0}\left(z^2-2\beta x_0\right)+\cdots.
\eeq
The leading order is suppressed by setting $\alpha=2/\sqrt{1+\s}$, in agreement with the preliminary analysis made above.
Finally (\ref{eq:schrod}) becomes the equation of a harmonic oscillator,
\beq\label{eq:OH}
\left(-\frac{\d^2}{ {\rm d} z^2}+\frac{\alpha}{4x_0}z^2\right)\tilde w(z)=\frac{\alpha\beta}{2}\tilde w(z),
\eeq
with energy $\alpha\beta/2$. 
The ground-state solution is\footnote{Let us remind that the harmonic oscillator
\beq
-y''+(x^2-\lambda)y=0,
\eeq
with $y(\pm\infty)=0$, has solutions in terms of Hermite polynomials, when $\lambda=2n+1$,
\beq
y_n(x)=\e^{-x^2/2}H_n(x).
\eeq
}
\beq
\tilde w(z)\propto\exp\left(-\frac{cz^2}{2}\right),
\eeq
that we cast in the equation above, in order to determine the constants $c$ and $\beta$ as,
\beq
c=\frac{1}{2}\sqrt{\frac{\alpha}{x_0}}=\frac{1}{\sqrt{2x_0}(1+\s)^{1/4}},\quad \beta=\frac{2c}{\alpha},
\eeq
so~(\ref{eq:OH}) simplifies into
\beq
\left(-\frac{\d^2}{ {\rm d} z^2}+c^2z^2\right)\tilde w(z)=c\,\tilde w(z).
\eeq
Finally
\beq\label{eq:Aprediction}
A=A_0-\frac{2}{\sqrt{1+\s}}\gamma^{1/2}+\frac{(1+\s)^{1/4}}{\sqrt{2x_0}}\gamma^{3/4}+\cdots.
\eeq
Coming back to the original variable $x$, we have
\beq\label{eq:predict-w}
w(x)\propto\exp\left(-\frac{c(x-x_0)^2}{2\gam^{3/4}}\right).
\eeq
The present analysis parallels that done for $b$ large in the $\s=1$ case,
where the scaling function is found to have finite support with an essential singularity~\cite{zeta2} (see the discussion below eq.~(3.23) therein).

Let us compare these predictions to the results of a numerical integration of~(\ref{eq:eqdif-first}).
Figure~\ref{fig:g+} depicts the solutions of this equation for various values of 
$\gam$, ranging from $\gam=0.4$ ($t\approx 15$) to $\gam\approx 8.5\, 10^{-6}$ ($t\approx 1.62\,10^{15}$), with $\s=1/2$ and $b=4$.
This figure demonstrates the very slow convergence of the finite-time solutions $g(x,t)$ to the stationary scaling function $g(x)$ given by~(\ref{eq:gstat}) with $A=A_0$.
The corresponding values of the amplitude $A$ are depicted in figure~\ref{fig:Anew+} (black dots).
These values are well predicted by~(\ref{eq:Aprediction}) which yields $A\approx 4.160-1.633\gamma^{1/2}+0.542\gamma^{3/4}$, with $\s=1/2$ and $b=4$ (red dashes).

\subsection{Complete equation~(\ref{eq:edpCond})}

As can be seen in figure~\ref{fig:Anew+}, the amplitude $A$ obtained by numerical integration of the master equation~(\ref{eq:master}) (yielding the complete equation~(\ref{eq:edpCond}) in the continuum limit) is different from the amplitude predicted for the simplified equation.
This discrepancy is due to the finite-time corrections induced by the term $t\dot g$ in the left side of~(\ref{eq:edpCond}).
The analysis of this situation is rather involved and is left for future work. 
As already mentioned earlier, for the $\s=1$ case, finite-time corrections are also present.
However these finite-time corrections are so small that there is no need to analyse their role.

\section{Discussion}

As mentioned in the introduction, the same condensing ZRP with rate~(\ref{eq:uksig}) was recently investigated in~\cite{gross}, with focus on coarsening in the condensed phase.
The scaling analysis of the single-site probability $f_k(t)$ given in this work
turns out to be incorrect.
In particular, the asymptotic scaling function $g(x)$ is not correctly predicted because $A$ is not taken equal to $A_0$, as imposed by the selection mechanism described in section~\ref{sec:scalingfctn};
for the analysis of the simplified equation~(\ref{eq:eqdif-first}) the time dependence of the amplitude $A$ in~(\ref{eq:eqdif-first}) is overlooked (it is conjectured instead to be linear in $b$ and depending on density, which does not hold);
the complete equation~(\ref{eq:edpCond}) is not derived.

\ack 
It is a pleasure to thank J M Luck for many enlightening discussions.

\pagebreak
\appendix
\setcounter{section}{0}

\section{Summary of formula}

We collect here some important formula for the ZRP under study, derived in the bulk of the paper, as well as the corresponding formula for the ZRP with rate $u_k=1+b/k$, derived in~\cite{zeta1,zeta2,cg2003}, for comparison.

\subsection*{$\bullet$ ZRP with rate $u_k=1+b/k$}

\subsubsection*{At criticality ($b>3$)}
\beq
\u=1+\eta_t,\quad\eta_t=A\eps^{b-2},\quad \eps=t^{-1/2}.
\eeq
\beq
f_k(t)= \left\{
\begin{array}{ll}
f_{k,\eq}(1+(v_0+k)\eta_t) & : k \textrm{ fixed}, t \textrm{ large}\vspace{4pt}
\\ 
f_{k,\eq}\,\g(k\eps) & : k \textrm{ and } t \textrm{ large}
\end{array}
\right.
\eeq
where $\g(x)$ satisfies
\beq\label{eq:gcproto}
\g^{\prime\prime }+\left(\frac{x}{2}-\frac{b}{x}\right) \g^{\prime}=0,
\eeq
and is explicitly given by~\cite{zeta2}
\beq\label{eq:gc_anal}
\g(x)=\frac{2^{-b}}{\Gamma (\frac{b+1}{2})}\int_{x}^{\infty
} {\rm d} y\,y^{b}\e^{-y^{2}/4}.
\eeq
The fall-off of $\g(x)$
for $x\gg 1$ is very fast: $\g(x)\sim \exp (-x^{2}/4)$,
hence $f_{k}(t)\sim \exp (-k^{2}/4t)$.
The amplitude $A$ is a function of $b$ alone and its explicit expression is known~\cite{zeta2}.

\subsubsection*{In the condensed phase}
\beq
\u=1+\eta_t,\quad\eta_t=A\eps,\quad \eps=t^{-1/2}.
\eeq
\beq
f_k(t)= \left\{
\begin{array}{ll}
f_{k,\eq}(1+(v_0+k)\eta_t) & : k \textrm{ fixed}, t \textrm{ large}\vspace{4pt}
\\ 
\eps^{2}g(k\eps) & : k \textrm{ and } t \textrm{ large}
\end{array}
\right.
\eeq
where $g(x)$ is solution of the equation
\beq\label{app:gcond}
g^{\prime\prime }+\left(\frac{x}{2}-A+\frac{b}{x}\right) g^{\prime}+\left( 1-\frac{b}{x^{2}}\right) g=0.
\eeq
The amplitude $A$ is again a function of $b$ alone.
Its explicit expression is only known for large values of $b$~\cite{zeta2}.

\subsection*{$\bullet$ ZRP with rate $u_k=1+b/k^\s$}

\subsubsection*{At criticality}
\beq
\u=1+\eta_t,\quad\eta_t\sim\e^{-t^{a}\varphi(0)},\quad \eps=t^{-1/(1+\s)}.
\eeq
\beq
f_k(t)= \left\{
\begin{array}{ll}
f_{k,\eq}(1+(v_0+k)\eta_t) & : k \textrm{ fixed}, t \textrm{ large}\vspace{4pt}
\\ 
f_{k,\eq}\,\g(k\eps,t) & : k \textrm{ and } t \textrm{ large}
\end{array}
\right.
\eeq
where $\g(x,t)$ satisfies
\beq
t\dot \g=t^{-a}\,\g''+\left(\frac{x}{1+\sigma}
-\frac{b}{x^\sigma}\right)\g',
\eeq
with $a=(1-\s)/(1+\s)$.

\subsubsection*{In the condensed phase}
\beq
\u=1+\eta_t,\quad\eta_t=A\eps^{\s},\quad \eps=t^{-1/(1+\s)}.
\eeq
\beq
f_k(t)= \left\{
\begin{array}{ll}
f_{k,\eq}(1+(v_0+k)\eta_t) & : k \textrm{ fixed}, t \textrm{ large}\vspace{4pt}
\\ 
\eps^{2}g(k\eps,t) & : k \textrm{ and } t \textrm{ large}
\end{array}
\right.
\eeq
where $g(x,t)$ is solution of 
\beq\
t\dot g=t^{-a}\,g''+\left(\frac{x}{1+\sigma}-A+\frac{b}{x^\sigma}\right)g'+\left(\frac{2}{1+\sigma}-\frac{\s b}{x^{1+\sigma}}\right)g.
\eeq


\end{document}